\documentclass{article}
\usepackage[paperwidth=7in, paperheight=10in, inner=0.7in, outer=0.7in, top=1in, bottom=1in]{geometry}
\usepackage{graphicx} 
\usepackage{amsmath,amsfonts}
\usepackage{fancyhdr}
\usepackage{amsmath, amssymb}
\usepackage{algorithm}
\usepackage{algpseudocode}
\usepackage{array}
\usepackage{caption}
\usepackage{pdflscape}
\usepackage{booktabs}
\usepackage{siunitx}
\usepackage{multirow}
\usepackage{imakeidx}
\usepackage{tocloft}
\usepackage{hyperref} 
\usepackage{titling}  
\usepackage[utf8]{inputenc}

\usepackage[
backend=biber,
style=nature
]{biblatex}
\begin{document}
\title{\textbf{Energy-efficient programmable integrated photonics via optimized Euler rotations}}

\date{}
\maketitle
\begin{center}

{\large
Pablo Martínez-Carrasco Romero,$^{a,*}$ 
Andrés Macho-Ortiz,$^{a,*}$ \\
José Roberto Rausell-Campo,$^{b}$ 
Francisco Javier Fraile-Peláez,$^{b}$ 
and José Capmany$^{a,c}$
}

\vspace{0.5cm}

\small
$^{a}$iTEAM Research Institute, Universitat Politècnica de València, Valencia, 46022, Spain \\

$^{b}$Dept. Teoría de la Señal y Comunicaciones, Universidad de Vigo E.I. Telecomunicación,\\
Campus Universitario, E-36202 Vigo (Pontevedra), Spain \\

$^{c}$iPronics, Programmable Photonics, S.L, Valencia, 46022, Spain \\

\vspace{0.3cm}

*Corresponding authors: 
pmarrom@iteam.upv.es, amachor@iteam.upv.es
\end{center}

\begin{center}
    \textbf{Abstract}
\end{center}

Programmable integrated photonics (PIP) has emerged as a powerful on-chip platform for optical signal processing and computing, enabling the implementation of reconfigurable N$\times$N unitary matrix transformations through meshes of tunable interferometers, which realize 2$\times$2 unitary matrices. However, the energy consumption associated with phase-shifter actuation is becoming a major limitation to the scalability of PIP platforms. Here, we introduce a geometric framework for energy optimization in PIP circuits by exploiting the representation of 2$\times$2 unitary matrices as concatenations of basic Euler rotations on the Bloch sphere. We show that equivalent implementations of the same N$\times$N unitary matrix (N $\geq$ 2) can exhibit markedly different energy costs depending on the rotation trajectories on the Bloch sphere implemented by each interferometer. Leveraging this insight, we identify minimum-energy configurations by systematically selecting the shortest rotation trajectories. We experimentally and numerically validate the proposed approach in diverse silicon PIP architectures, including feedforward and multipurpose hexagonal meshes, neural network accelerators, and photonic quantum-gate implementations. These results establish a general route toward more energy-efficient large-scale PIP processors for classical and quantum signal processing and computing applications.\\


\clearpage
\newpage

{\Large \textbf{1 Introduction}}\\
\begin{refsection}

Large-scale programmable integrated photonic (PIP) circuits are emerging as a powerful on-chip platform for optical signal processing and computing, driven by the increasing limitations of digital electronic systems \cite{intro1,intro2} and leveraging the unique advantages of photonics, including high bandwidth, low latency, and inherent parallelism \cite{intro3,intro4}. These capabilities enable a broad range of applications, ranging from quantum information processing \cite{intro5,intro6,intro7,intro8} and optical neural networks \cite{intro9,intro10,intro11} to advanced signal processing \cite{intro12,intro13} and reconfigurable optical switching in data centers \cite{intro14,intro15}. Among these, hardware-acceleration tasks are particularly compelling, as they rely on the implementation of high-dimensional unitary matrix transformations for matrix-vector multiplications in neuromorphic and quantum computing systems \cite{intro16,intro17}. In this context, PIP enables efficient matrix signal processing, leveraging its physical parallelism to address fundamental data-movement and scalability bottlenecks in conventional digital electronic architectures \cite{intro1,intro2,intro3,intro4}. \\

As shown in Fig. \ref{fig1}, these PIP processors implement reconfigurable N×N unitary matrices by electrically controlling light propagation through meshes of interconnected optical waveguides \cite{intro18,intro19,singlechip}. Their functionality is built upon tunable interferometric building blocks (BBs), which constitute the elementary technological units or gates of PIP \cite{lpr23}. Each BB implements a reconfigurable 2×2 unitary matrix \cite{lpr21}, enabling the realization of arbitrary high-dimensional unitary matrices through the interconnection of these BBs \cite{reck,clements}. From a geometric standpoint, each 2×2 unitary operation implemented by the BBs can be regarded as a distinct rotation on the Bloch sphere between two points (or states). Such a rotation can be factorized as a sequence of basic rotations around its Cartesian axes, with distinct equivalent combinations yielding the same 2×2 transformation \cite{lpr21}.\\

Recent advances in foundry-scale silicon photonics have enabled these architectures to scale up to hundreds and even thousands of BBs, significantly enhancing their signal processing capabilities \cite{scale1,scale2}. However, as the number of BBs increases, the cumulative physical actuation required to control and stabilize the circuit grows accordingly, making energy consumption a critical bottleneck for large-scale integration due to increasing control overhead, energy-density limitations, and system complexity \cite{scale3,scale4}. In this context, further scaling is fundamentally constrained by the phase shifter, the key device enabling reconfigurability and the dominant contributor to energy consumption at both the BB and photonic processor levels.

While improving phase-shifter efficiency remains an active area of research \cite{scale5}, alternative approaches have recently focused on reducing the overall phase requirements of the circuit through system-level strategies such as pruning of less relevant components, as well as phase redistribution strategies within the mesh \cite{hamerly1,hamerly2,pruning}. Previous works have shown that not all BBs contribute equally to a given unitary transformation, revealing an intrinsic non-uniformity in their role and enabling sparsity-driven pruning strategies that reduce both energy consumption and circuit complexity \cite{pruning}. Complementary efforts, inspired by information-theoretic principles, have proposed alternative phase-allocation strategies within a given circuit architecture, redistributing phase across the mesh and reducing the average phase required from individual phase shifters \cite{hamerly2}. Despite these advances, such approaches focus on how unitary transformations are distributed across the circuit, rather than on how they are physically implemented. \\

Therefore, a more fundamental degree of freedom remains largely unexplored: the mathematical degeneracy underlying the physical implementation of individual 2×2 unitary transformations, arising from the existence of distinct equivalent decompositions into basic rotations around the Cartesian axes of the Bloch sphere (Fig. \ref{fig1}c). Remarkably, these equivalent implementations can exhibit markedly different energy costs.

\begin{figure}[ht]
\centering\includegraphics[width=0.99\linewidth]{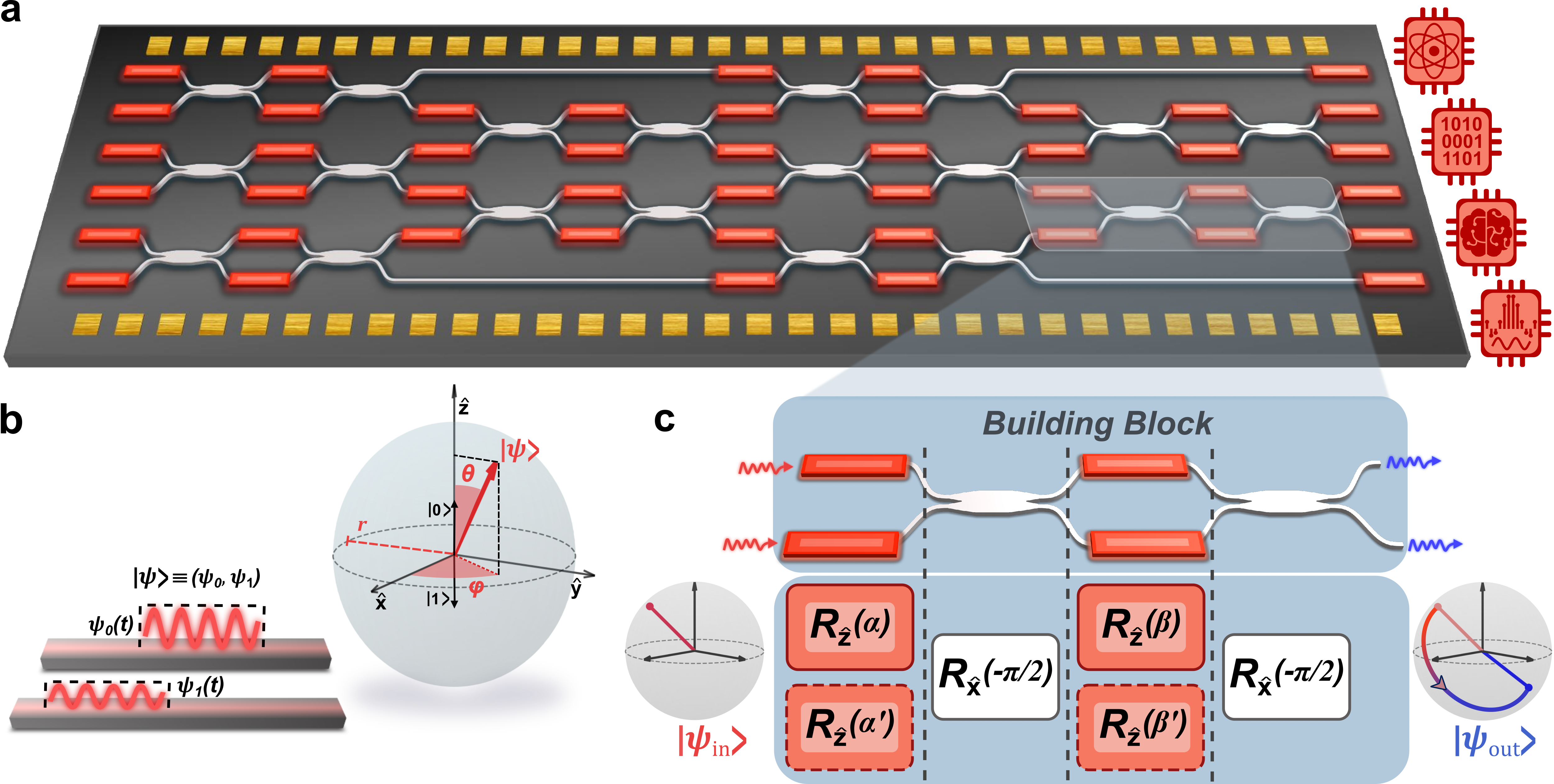} 
\caption{ \textbf{Programmable integrated photonic (PIP) circuits for unitary matrix transformations.} \textbf{a} Schematic of a PIP circuit implementing reconfigurable 6×6 unitary matrix transformations. \textbf{b}  Representation on the Bloch sphere of the optical field propagating through two parallel waveguides (parallel-waveguide section), forming an effective two-mode system. \textbf{c}  Schematic of a tunable interferometric building block (BB) composed of beam splitters and phase shifters. The BB implements a reconfigurable 2×2 unitary transformation between two adjacent modes of the PIP mesh, geometrically represented as a rotation on the Bloch sphere between the input and output states. Equivalent implementations of the same transformation can be obtained through different sets of rotation angles associated with the same sequence of Euler rotations on the Bloch sphere.}
\label{fig1}
\end{figure}

In this work, we exploit this previously unaddressed degree of freedom by introducing a geometric framework to analyze and minimize energy consumption at the level of the fundamental BB. By interpreting the classical optical field transformations within each BB as basic rotations around the Cartesian axes of the Bloch sphere, we show that equivalent 2×2 unitary transformations can be implemented through different rotation paths, each associated with a distinct total phase accumulation and, consequently, a different energy cost. Leveraging this insight, we identify minimum-energy implementations by systematically selecting the shortest rotation trajectories, providing a physically grounded and scalable route toward energy-efficient PIP circuits. In addition, since quantum optical field transformations within each BB are governed by the same rotation operations \cite{intro33,intro34}, this approach establishes a general principle for minimizing energy consumption in large-scale PIP processors for both classical and quantum applications.
\clearpage
\newpage

{\Large \textbf{2 Results}}\\\\
{\textbf{2.1 Theoretical framework}}\\

Since BBs and their 2×2 unitary transformations are the fundamental elements that enable scaling to large N×N unitary matrices, their accurate and energy-efficient implementation is critical for the performance and scalability of PIP meshes. Accordingly, we first focus on the realization and optimization of unitary transformations at the BB level (N $=$ 2) and then extend the approach to higher-dimensional unitary matrices (N $>$ 2). To optimize energy consumption, we adopt a geometric framework that captures the underlying wave transformations and enables systematic energy optimization at both the BB level and across the entire PIP mesh.\\

We therefore begin by analyzing the BB, focusing on the representation of the optical fields at its input and output (Fig. \ref{fig1}b), and on the transformation relating them (Fig. \ref{fig1}c). In particular, we consider a two-mode system in which the electromagnetic fields propagate through a pair of optical waveguides, forming an effective two-dimensional state. This formulation provides a natural framework to describe the action of the BB as a transformation between input and output states, establishing the basis for its subsequent geometric interpretation.

As shown in Fig. \ref{fig1}, the input (and output) state is composed of two optical wave packets (or complex envelopes) $\psi_{0}$ and $\psi_{1}$ that can be geometrically represented as a three-dimensional point with spherical coordinates $(r,\theta,\varphi)$, where $r=\sqrt{|\psi_{0}|^{2}+|\psi_{1}|^{2}}$ is the radius, $\theta = 2\arctan(|\psi_{1}|/|\psi_{0}|)$ is the elevation angle, and $\varphi=\mathrm{Arg}(\psi_{1})-\mathrm{Arg}(\psi_{0})$ is the azimuthal angle. In analogy with quantum computing \cite{intro17,lpr23} and analog programmable-photonic computing \cite{lpr21}, this representation can be interpreted as a classical state  $|\psi\rangle = \psi_{0}|0\rangle+\psi_{1}|1\rangle \equiv r(\cos{(\theta/2)}|0\rangle+e^{i\varphi}\sin{(\theta/2)}|1\rangle)$ located on the surface of the Bloch sphere, with the standard states  $|0\rangle$ and  $|1\rangle$ describing the fundamental modes of the two optical waveguides. Here, the relative phase of the two-mode optical field is encoded in the azimuthal angle. However, the global phase is not included in this geometric representation since, in most current PIP applications, including direct and interferometric detection schemes, it does not affect the detected signals or the implemented functionality \cite{intro17}. Accordingly, the global phase is omitted throughout the optimization framework presented in this work. The implications of global phase in the context of energy optimization are addressed later in the Discussion section.\\

Despite the variety of BB designs proposed for scalable unitary transformations \cite{hamerly1}, they are all based on 50:50 beam splitters combined with phase shifters and exhibit fundamentally equivalent behavior. As established in previous works \cite{lpr23,lpr21}, each BB implements a 2×2 unitary transformation that can be geometrically interpreted as a rotation on the Bloch sphere relating the input and output optical states. Here, 50:50 beam splitters, typically implemented using symmetric directional couplers or multimode interferometers (MMIs), perform fixed rotations $ R_{\hat{\mathrm{\textbf{x}}}}(\pm\pi/2)$ around the x-axis of the Bloch sphere, whereas phase shifters implement tunable rotations  $R_{\hat{\mathrm{\textbf{z}}}}(\alpha)$ around the z-axis, with a reconfigurable rotation angle $\alpha$.

Among the various BB implementations reported in the literature, we adopt here a generalized architecture composed of two 50:50 beam splitters and parallel-waveguide sections incorporating phase shifters in both arms (Fig. \ref{fig2}). This choice is motivated by two key features. First, the resulting transfer matrix guarantees complete access to the unitary rotation space of the Bloch sphere \cite{lpr21}. Second, as shown below, the simultaneous control of both waveguides introduces an additional degree of freedom that enables lower-energy implementations of equivalent unitary transformations compared with conventional BB designs in which each parallel-waveguide section incorporates only a single phase shifter.

\begin{figure}[ht]
\centering\includegraphics[width=0.85\linewidth]{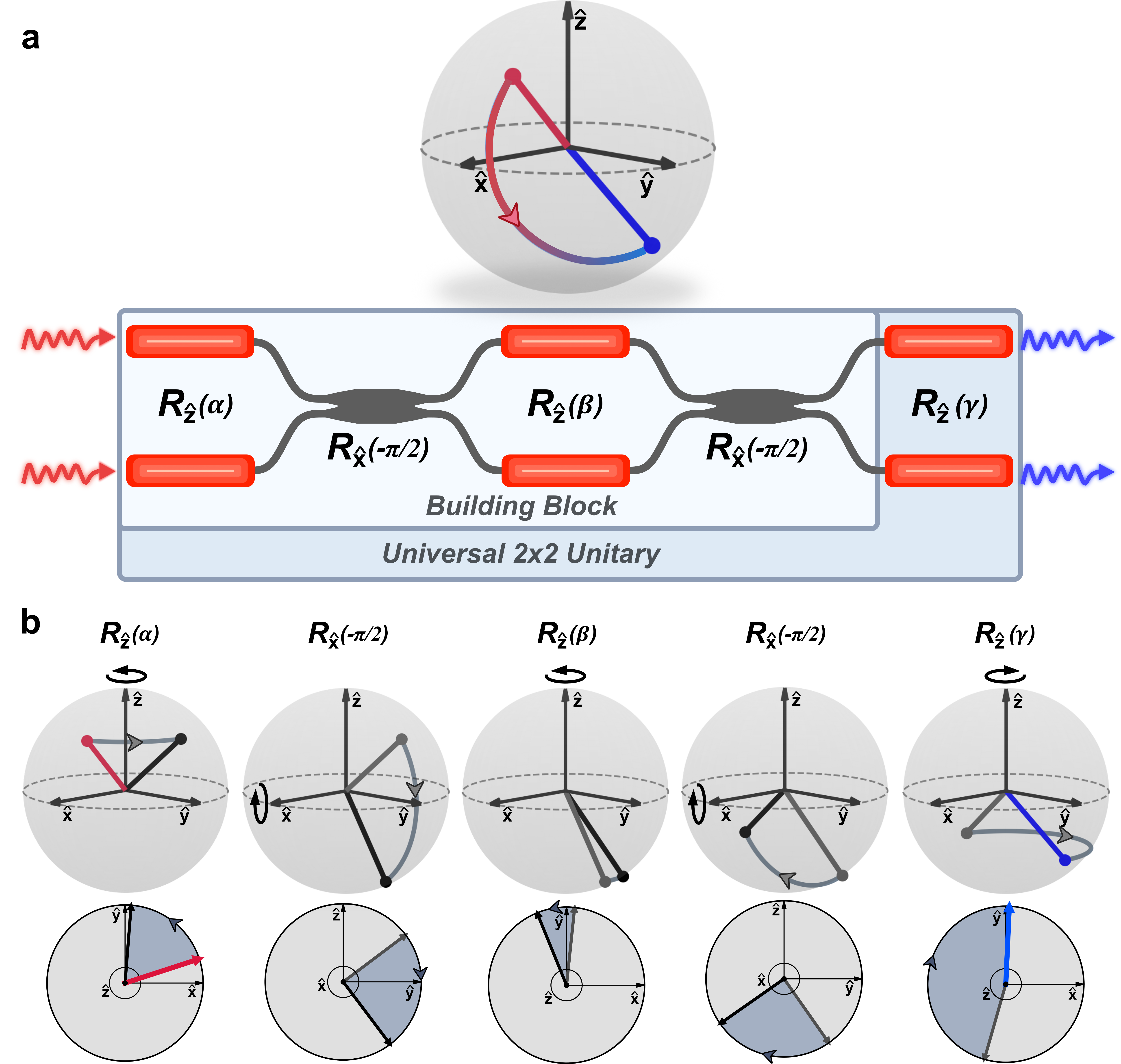} 
\caption{ \textbf{Geometrical interpretation of universal 2×2 unitary matrix transformations in a PIP circuit.} \textbf{a} Schematic of a PIP processor implementing arbitrary 2×2 unitary matrix transformations. The circuit transforms an arbitrary input state (red) into an arbitrary output state (blue) on the Bloch sphere through a rotation connecting both states. This transformation can be decomposed into five independent basic rotations around the Cartesian axes of the Bloch sphere. \textbf{b} Geometrical representation of the basic rotations composing the complete transformation, shown both as three-dimensional Bloch-sphere trajectories and as two-dimensional projections along the corresponding rotation axis ($\hat{\mathrm{\textbf{z}}}$ or $\hat{\mathrm{\textbf{x}}}$, depending on the rotation).}
\label{fig2}
\end{figure}

As depicted in Fig. \ref{fig2}a, by cascading the basic rotations that compose the transfer matrix ($T_{\mathrm{BB}}$) of the BB, $T_{\mathrm{BB}}(\alpha,\beta) = R_{\hat{\mathrm{\textbf{x}}}}(-\pi/2)R_{\hat{\mathrm{\textbf{z}}}}(\beta)R_{\hat{\mathrm{\textbf{x}}}}(-\pi/2)R_{\hat{\mathrm{\textbf{z}}}}(\alpha)$, together with an additional parallel-waveguide section at the output implementing the transformation$R_{\hat{\mathrm{\textbf{z}}}}(\gamma)$, Euler’s rotation theorem guarantees that the resulting PIP circuit, with transfer matrix:

\begin{equation}
    T_{2\times2}(\alpha,\beta,\gamma) = R_{\hat{\mathrm{\textbf{z}}}}(\gamma)T_{\mathrm{BB}}(\alpha,\beta) =i
    \begin{pmatrix} 
     -e^{-i(\alpha+\gamma)/2}\sin\left(\dfrac{\beta}{2}\right) & e^{i(\alpha-\gamma)/2}\cos\left(\dfrac{\beta}{2}\right) \\
     e^{-i(\alpha-\gamma)/2}\cos\left(\dfrac{\beta}{2}\right) & e^{i(\alpha+\gamma)/2}\sin\left(\dfrac{\beta}{2}\right) \\
    \end{pmatrix},
\end{equation}

can implement any 2×2 unitary transformation \cite{lpr21}. This universal unitary matrix can be mapped onto any target unitary matrix to extract the corresponding set of equivalent rotation angles. Figure \ref{fig2}b illustrates the geometric representation on the Bloch sphere of this decomposition into basic Euler rotations for a representative set of angles $(\alpha,\beta,\gamma)$. In this system, the total phase accumulation $|\alpha|+|\beta|+|\gamma|$ emerges as a physically meaningful metric of the phase-shifter actuation required to implement the target transformation and, consequently, of its associated energy cost across different phase-shifter technologies (Supplementary Section \ref{supp1}).

Importantly, the mapping between a target unitary transformation and its corresponding set of Euler rotation angles \textit{is not unique}, owing to an underlying mathematical degeneracy. Multiple equivalent combinations of $(\alpha,\beta,\gamma)$ may generate the same 2×2 unitary transformation while following different rotation trajectories on the Bloch sphere, thereby connecting the initial and final optical state through distinct geometric paths. As a consequence, these physically equivalent implementations can exhibit markedly different total phase accumulations $|\alpha|+|\beta|+|\gamma|$ and, therefore, different energy consumptions in the unitary processor. Even when restricting the rotation angles to the fundamental interval $|\alpha|+|\beta|+|\gamma|\in[0,2\pi)$, two distinct sets of angles $(\alpha,\beta,\gamma)$ remain for a given 2×2 unitary transformation, one of which systematically requires a larger total phase accumulation, thus leading to a higher energy consumption (see Supplementary Section 2 for more details). Within this framework, the optimal implementation corresponds to the set of Euler angles that minimizes the total phase accumulation, $|\alpha|+|\beta|+|\gamma|$, and hence the associated energy cost.\\

While the proposed optimization method is general and applicable to a broad range of programmable phase-shifter technologies (see Discussion), we focus here on thermo-optic phase shifters as a representative case study owing to their widespread use in current PIP platforms \cite{intro17, singlechip}.
In thermo-optic technology, the induced phase shift arises from a temperature-dependent refractive-index variation $\Delta n(T)$ over a modulation length $L$, yielding an accumulated phase proportional to $L\Delta n(T)$. Since the refractive-index variation induced by heating is inherently positive and temperature can only increase, the physically accessible phase shifts are constrained to positive values. Consequently, this physical constraint raises the question of whether both clockwise and counterclockwise $R_{\hat{\mathrm{\textbf{z}}}}$ rotations on the Bloch sphere remain physically accessible within the BB. 

The answer to this question depends on the specific implementation of the $R_{\hat{\mathrm{\textbf{z}}}}$ rotations within the BB. When these rotations are realized using two independently actuated phase shifters in the parallel-waveguide sections of the BB (Fig. 3a), the resulting transfer matrix corresponds to a tunable $R_{\hat{\mathrm{\textbf{z}}}}(\alpha)$ rotation with $\alpha = \alpha_{2}-\alpha_{1}$, where $\alpha_{1}$ and $\alpha_{2}$ denote the phase shifts applied to the upper and lower waveguides, respectively (see Supplementary Section \ref{supp2}). As a result, both positive and negative rotation angles — and hence both clockwise and counterclockwise rotations on the Bloch sphere — remain physically accessible. In contrast, a single-phase-shifter implementation (e.g. in the upper waveguide, see Fig. \ref{fig3}b) only enables rotations of one sign (clockwise rotations), thus restricting the accessible rotation direction on the Bloch sphere.

\clearpage

\begin{figure}[ht]
\centering\includegraphics[width=0.83\linewidth]{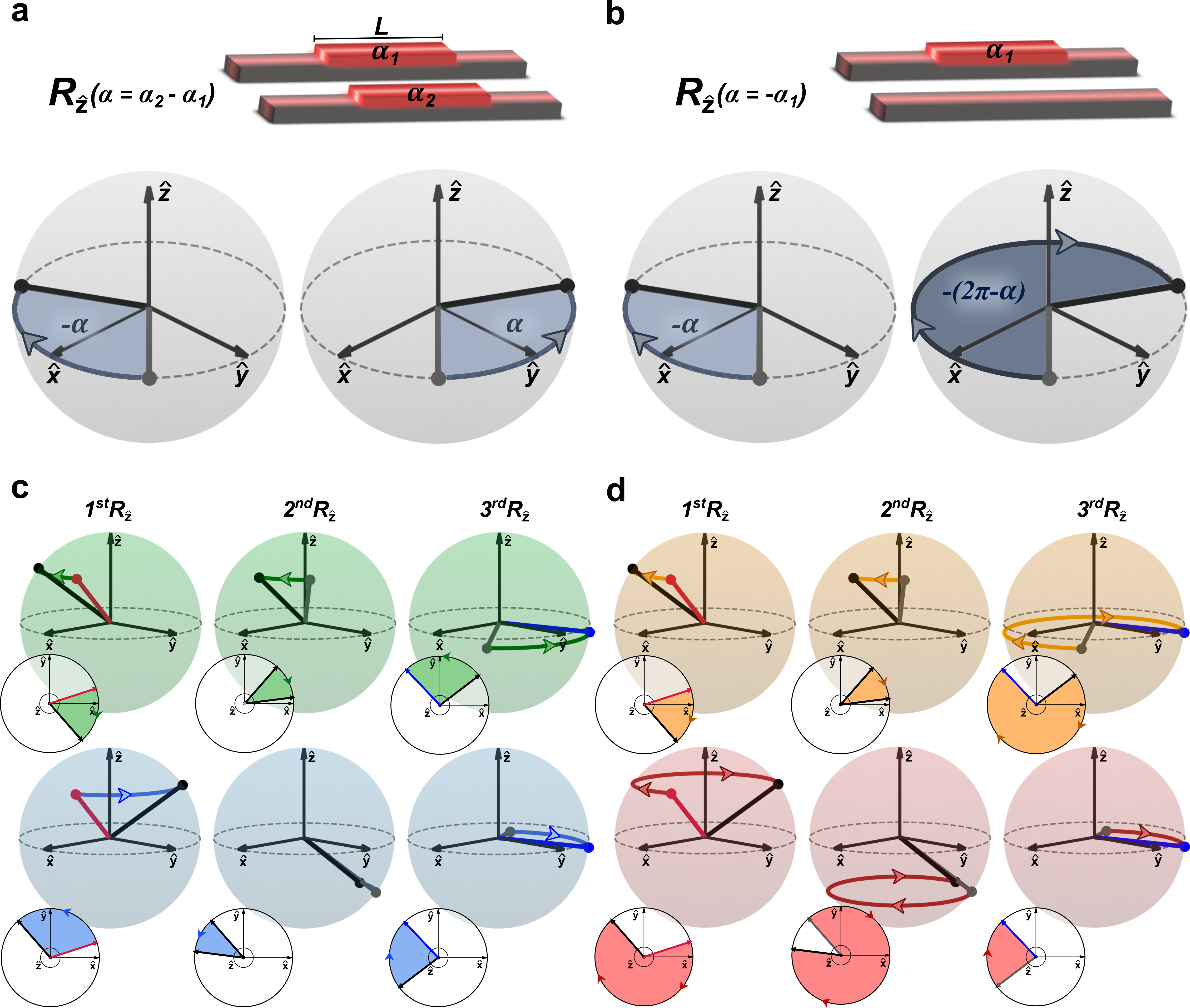} 
\caption{ \textbf{Comparison between dual- and single-phase-shifter implementations of tunable $\mathbf{R_{\hat{\mathrm{z}}}}$ rotations in a parallel-waveguide section.} \textbf{a} Parallel-waveguide section with two independently actuated thermo-optic phase shifters, enabling both clockwise and counterclockwise $R_{\hat{\mathrm{\textbf{z}}}}$ rotations on the Bloch sphere through the relative phase difference $\alpha = \alpha_{2}-\alpha_{1}$.  \textbf{b}  Parallel-waveguide section with a single thermo-optic phase shifter, restricting the physically accessible rotation direction and forcing the implementation of complementary rotations $2\pi-\alpha$ for inaccessible rotation signs. \textbf{c} Two degenerate sets of Euler angles $(\alpha,\beta,\gamma)$ implementing the same 2×2 unitary matrix using independently actuated phase shifters in each parallel-waveguide section associated with a tunable $R_{\hat{\mathrm{\textbf{z}}}}$ rotation. The minimum- and maximum-energy trajectories are highlighted in green and blue, respectively. \textbf{d} The same unitary transformation implemented using a single active phase shifter in each parallel-waveguide section associated with a tunable $R_{\hat{\mathrm{\textbf{z}}}}$ rotation. The restriction to a single rotation direction forces the use of complementary rotations, increasing the total phase accumulation and the associated energy consumption. The corresponding minimum- and maximum-energy trajectories are shown in orange and red, respectively.}
\label{fig3}
\end{figure}

As a consequence, once a target 2×2 unitary transformation is decomposed into its corresponding sets of Euler angles $(\alpha,\beta,\gamma)$, the minimum-energy solution can be directly implemented when two independently actuated phase shifters are employed in each parallel-waveguide section of the BB. In contrast, single-phase-shifter architectures require the use of complementary rotations for solutions involving inaccessible rotation signs, thus increasing the total phase accumulation and the associated energy cost for the same unitary transformation.
The combined impact of angular degeneracy and physical implementation constraints is illustrated in Figs. \ref{fig3}c and \ref{fig3}d for a representative 2×2 unitary transformation. 

For clarity, only the tunable $R_{\hat{\mathrm{\textbf{z}}}}$ rotations are represented, whereas the fixed $R_{\hat{\mathrm{\textbf{x}}}}$ rotations are omitted since they do not contribute to the energy consumption. In Fig. \ref{fig3}c, the two degenerate sets of Euler angles associated with the same unitary transformation are implemented using two independently actuated phase shifters for each $R_{\hat{\mathrm{\textbf{z}}}}$ operation.  The corresponding minimum- and maximum-energy trajectories are highlighted in green and blue, respectively. Conversely, Fig. \ref{fig3}d depicts the same sets of angles implemented using a single active phase shifter per $R_{\hat{\mathrm{\textbf{z}}}}$ transformation. In this case, the restriction to a single accessible rotation direction forces the use of complementary rotations, increasing the total accumulated phase and the associated energy consumption for the same unitary transformation.

Having established the optimization framework for 2×2 unitary transformations under realistic physical constraints, the proposed method naturally extends to arbitrary N×N unitary matrices. In feedforward architectures such as Reck or Clements \cite{reck,clements}, the target unitary matrix is decomposed into a sequence of elementary two-level unitary matrices $T_{m,n}(\alpha, \beta)$ through successive nullification of off-diagonal elements. Each elementary operation acts on a pair of waveguides $(m,n)$ and corresponds to an embedded 2×2 unitary transformation $\tilde{T}_{m,n}(\alpha, \beta)$ that can be decomposed into the same sequence of Euler rotations implemented by the BB, $\tilde{T}_{m,n}(\alpha, \beta)\equiv R_{\hat{\mathrm{\textbf{x}}}}(-\pi/2)R_{\hat{\mathrm{\textbf{z}}}}(\beta)R_{\hat{\mathrm{\textbf{x}}}}(-\pi/2)R_{\hat{\mathrm{\textbf{z}}}}(\alpha)$, enabling independent energy optimization of each BB through minimization of the total phase accumulation (see Supplementary Section \ref{supp3} for further details).

Remarkably, although the present framework has been introduced using classical optical fields, the same formalism directly applies to quantum photonic systems. In particular, both classical and quantum optical transformations implemented within the BB are governed by the same unitary rotation operators and the Euler decomposition theorem on the Bloch sphere \cite{lpr21}. Consequently, the proposed optimization strategy establishes a general principle for minimizing energy consumption in large-scale PIP processors for both classical and quantum photonic applications.\\

{\textbf{2.2 Experimental and numerical validation}}\\

To experimentally validate the proposed optimization framework, we performed a sequence of demonstrations addressing progressively increasing levels of complexity and scalability. We first implemented a universal 2×2 unitary processor as a proof of concept to experimentally verify the impact of Euler-decomposition degeneracy and different phase-shifter implementations of $R_{\hat{\mathrm{\textbf{z}}}}$ rotations on energy consumption. We then scale the approach to high-dimensional PIP circuits by implementing arbitrary unitary transformations in a multipurpose hexagonal mesh architecture. Finally, we investigated the implications of the proposed optimization strategy in practical optical-computing scenarios through numerical demonstrations involving feedforward neural networks and photonic quantum-gate implementations.

We first implemented the universal 2×2 unitary processor shown in Fig. \ref{fig4}a in a silicon PIP circuit fabricated on a silicon-on-insulator (SOI) platform (see Methods). To experimentally validate the proposed optimization framework, we implemented 2000 random 2×2 unitary transformations and quantified the corresponding power consumption for four different implementation cases arising from the combination of: $(i)$ the two degenerate sets of Euler angles associated with each unitary transformation, and $(ii)$ the use of either one or two independently actuated phase shifters in each tunable $R_{\hat{\mathrm{\textbf{z}}}}$ rotation. The resulting power-consumption histograms are shown in Fig. \ref{fig4}b using the same color coding introduced in Fig. \ref{fig3}. Green and blue correspond to the lowest- and highest-power-consumption implementations associated with the two degenerate sets of Euler angles obtained when both phase shifters are independently actuated for each tunable $R_{\hat{\mathrm{\textbf{z}}}}$ rotation, directly revealing the impact of Euler-decomposition degeneracy on power consumption. Orange and red show the corresponding lowest- and highest-power-consumptions obtained using a single thermo-optic phase shifter for each $R_{\hat{\mathrm{\textbf{z}}}}$ transformation, where the restriction to a single rotation direction increases the total phase accumulation and the associated power consumption.

\begin{figure}[ht]
\centering\includegraphics[width=0.99\linewidth]{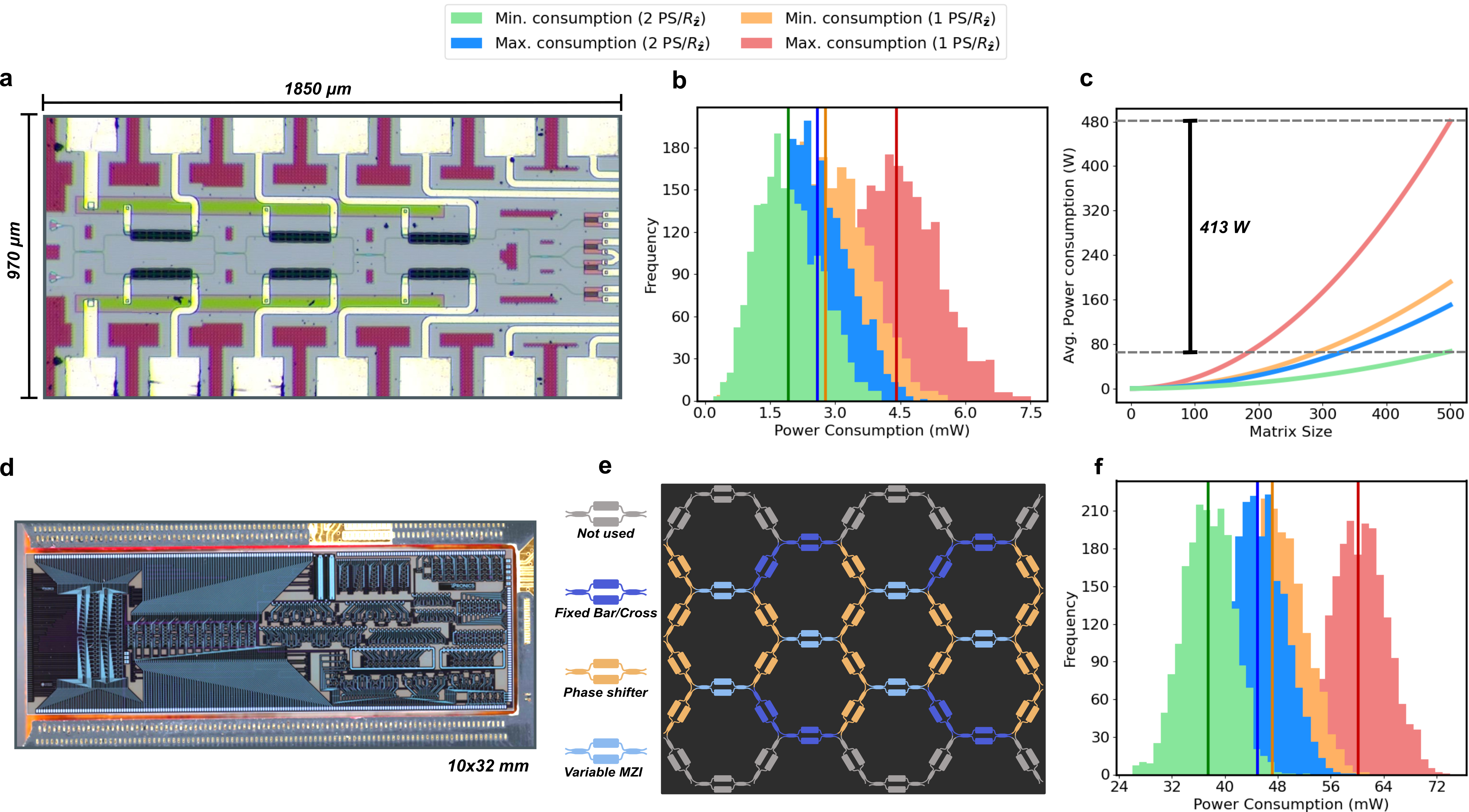} 
\caption{ \textbf{Experimental validation and scalability analysis of the proposed energy optimization method across diverse PIP architectures.} \textbf{a} Micrograph of the silicon photonic chip implementing a universal 2×2 unitary processor, including the electrical interconnects and integrated balanced photodetection system.  \textbf{b}  Experimental power-consumption histograms obtained from the implementation of 2000 random 2×2 unitary transformations for different combinations of Euler-decomposition degeneracy and phase-shifter implementations of $R_{\hat{\mathrm{\textbf{z}}}}$ rotations. Green and blue correspond to the minimum- and maximum-power-consumption sets of Euler angles implemented using two active phase shifters per tunable $R_{\hat{\mathrm{\textbf{z}}}}$ rotation, whereas orange and red show the corresponding minimum- and maximum-power- consumptions when using a single active phase shifter per $R_{\hat{\mathrm{\textbf{z}}}}$ transformation. \textbf{c} Numerically estimated average power consumption for unitary matrix transformations with dimensions ranging from 2×2 to 500×500. \textbf{d} Micrograph of the multipurpose PIP processor based on a hexagonal mesh topology (Smartlight, iPronics Programmable Photonics). \textbf{e} Mapping of a feedforward unitary architecture onto the PIP hexagonal mesh for the implementation of 4×4 unitary matrices. \textbf{f} Experimental power-consumption histograms obtained from the implementation of 2000 random 4×4 unitary transformations in the PIP hexagonal mesh using one or two independently actuated phase shifters per tunable Mach-Zehnder interferometer (MZI).}
\label{fig4}
\end{figure}

We further investigated the scalability of the proposed optimization framework by extending the analysis to high-dimensional unitary transformations. Assuming identical power consumption per phase shifter, we numerically estimated the average power required to implement unitary matrices with dimensions ranging from 2×2 to 500×500. As shown in Fig. \ref{fig4}c, the total power consumption increases rapidly with matrix dimensionality when non-optimal $R_{\hat{\mathrm{\textbf{z}}}}$  rotation implementations are employed, revealing the growing impact of BB-level energy optimization in large-scale PIP circuits. For 500×500 unitary matrices, the optimized implementation reduces the total power consumption by approximately 413 W compared to the highest power-consumption configuration.

We next experimentally validated the proposed optimization framework in a multipurpose PIP processor based on a hexagonal mesh topology (Smartlight, iPronics Programmable Photonics, see Fig. \ref{fig4}d and Methods). Following previously reported mappings of feedforward unitary architectures onto PIP meshes \cite{singlechip,reck,clements,scale3}, we implemented 2000 random 4×4 unitary transformations (see Fig. \ref{fig4}e) using one or two independently actuated phase shifters per tunable Mach-Zehnder interferometer (MZI) of the mesh. The resulting power-consumption histograms are shown in Fig. \ref{fig4}f using the same color coding introduced in Fig. \ref{fig3}. Although the programmable hexagonal mesh exhibits a higher overall power consumption than dedicated feedforward architectures owing to the additional MZIs required for light routing (see Supplementary Section \ref{supp4}), the same optimization trends are preserved, experimentally confirming the applicability of the proposed framework to scalable multipurpose PIP processors.\\

Finally, we numerically investigated the implications of the proposed energy optimization method in practical optical computing scenarios, considering representative applications in feedforward neural networks and photonic quantum gates. 

For the neural-network application, we considered a three-hidden-layer feedforward architecture trained on the MNIST classification task \cite{mnist}. To reduce the input dimensionality, the original 28×28 pixel images were compressed into a 4×4 representation by retaining the 16 lowest-frequency Fourier components. The resulting photonic weight matrices were simulated using a Clements architecture with thermo-optic phase shifters, considering hidden-layer sizes ranging from 32 to 256 neurons (Fig. \ref{fig5}a). The estimated power consumption for the different implementation strategies is shown in Fig. \ref{fig5}b. The results reveal that the impact of Euler-decomposition degeneracy and the implementation of tunable $R_{\hat{\mathrm{\textbf{z}}}}$ rotations using one or two active phase shifters become increasingly pronounced as the neural-network dimensionality increases. In particular, the optimized implementations reduce the total power consumption by factors ranging from approximately ×5.5 to ×8.2 compared to the highest-power-consumption configurations, with the largest energy-savings obtained for the largest hidden layers.

We further investigated the impact of Euler-decomposition degeneracy and one- versus two-phase-shifter implementations of $R_{\hat{\mathrm{\textbf{z}}}}$ rotations on the optical implementation of representative two-qubit gates (described by 4×4 unitary matrices), including Controlled-Hadamard (CH), FSim, and RZX gates \cite{gate1,gate2}. Arbitrary 4×4 unitary matrices were synthesized using the KAK decomposition \cite{kak}, as illustrated in Fig. \ref{fig5}c. The estimated power consumption for the different quantum-gate implementations is shown in Fig. \ref{fig5}d, revealing power-consumption reductions ranging from approximately ×1.4 to ×2.8 depending on the implemented quantum gate. These results demonstrate that the proposed optimization framework can also significantly reduce the power consumption of photonic quantum processors.
\clearpage

\begin{figure}[ht]
\centering\includegraphics[width=0.99\linewidth]{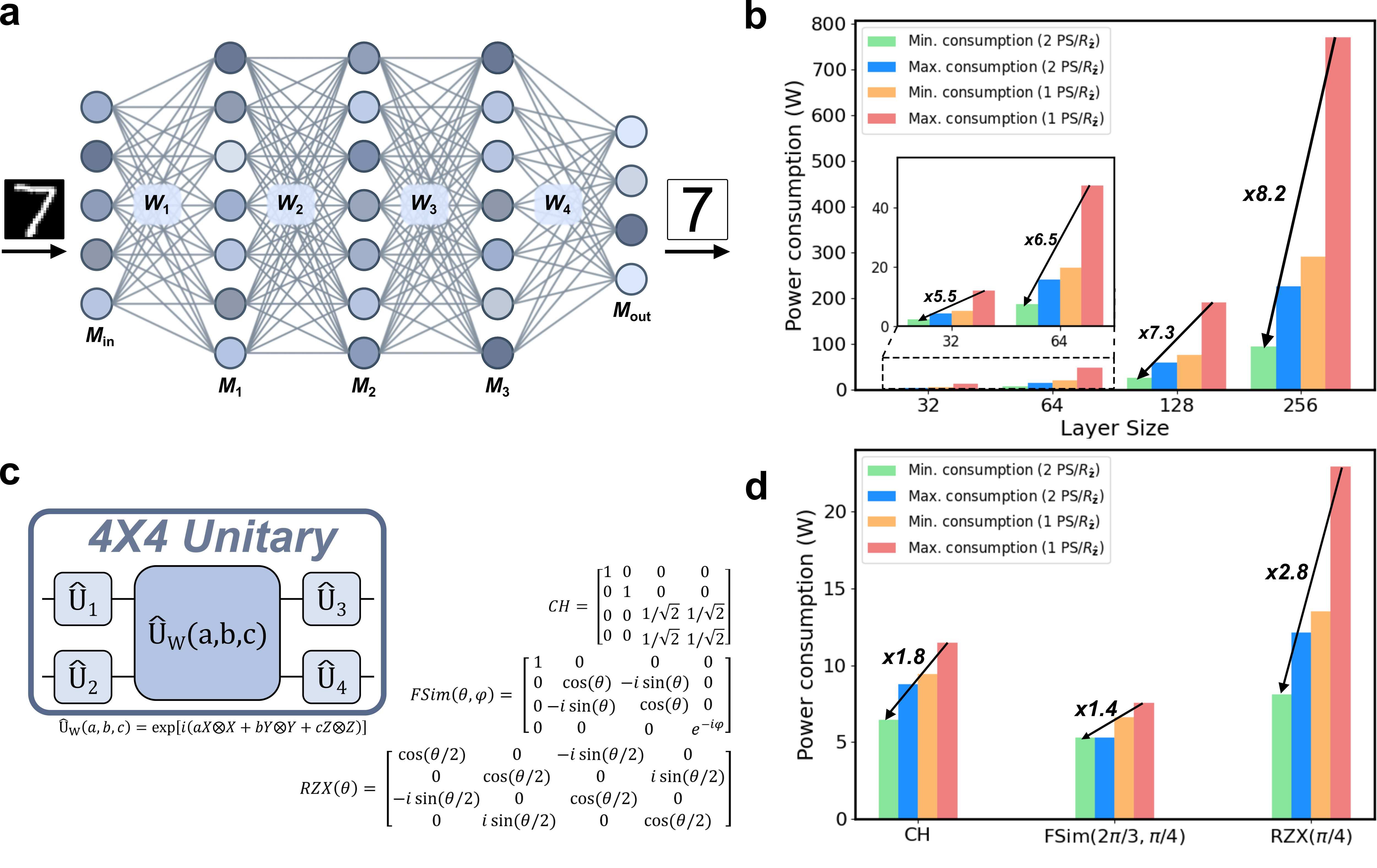} 
\caption{ \textbf{Power-consumption optimization in optical-computing applications.} \textbf{a} Schematic of the feedforward neural-network architecture used for the numerical analysis of power consumption. The numbers of neurons in the input and output layers, $M_{\mathrm{in}}$ and $M_{\mathrm{out}}$, are fixed to 16 and 10, respectively, whereas the number of neurons in the hidden layers, $W_{k}$ $(k=1,2,3)$, varies from 32 to 256. $W_{p}$ denotes the photonic weight matrix connecting the p-th and $p$-th layers. \textbf{b}  Estimated power consumption as a function of hidden-layer size for different implementations arising from Euler-decomposition degeneracy and one- versus two-phase-shifter implementations of tunable $R_{\hat{\mathrm{\textbf{z}}}}$ rotations. \textbf{c} Circuit representation of a two-qubit gate (arbitrary 4×4 unitary matrices) synthesized using the KAK decomposition ($\hat{U}_{1,2,3,4}$ are local single-qubit operations and $\hat{U}_{W}$ is a non-local two-qubit operation describing entanglement \cite{kak}), together with representative numerical examples. \textbf{d} Estimated power consumption for different photonic quantum-gate implementations under different Euler-decomposition and $R_{\hat{\mathrm{\textbf{z}}}}$-implementation strategies.}
\label{fig5}
\end{figure}

{\Large \textbf{3 Discussion}}\\

Overall, we have introduced an energy optimization framework for PIP circuits based on the geometric interpretation of unitary transformations on the Bloch sphere. By exploiting the mathematical degeneracy underlying the Euler decomposition of 2×2 unitary matrices, we demonstrated that equivalent factorizations of the same unitary matrix into basic Euler rotations can exhibit markedly different energy consumptions, with minimum-energy implementations naturally emerging from the shortest rotation trajectories on the Bloch sphere. The proposed approach was experimentally and numerically validated across a broad range of classical and quantum photonic signal processing and computing platforms, including feedforward processors, multipurpose PIP meshes, neural network accelerators, and photonic quantum-gate implementations.

The proposed optimization framework has important implications for the scalability of PIP circuits. As PIP processors continue to increase in size and functionality, the cumulative energy consumption associated with the actuation and stabilization of thousands of phase shifters is emerging as a major limitation to further scaling \cite{scale1,scale2}. In this context, the results obtained for high-dimensional unitary transformations indicate that relatively small energy differences at the BB level can accumulate into substantial system-level power savings as the matrix dimensionality increases, reaching approximately 30, 100 and 400 W for optimized 128×128, 256×256, and 512×512 unitary matrices, respectively, compared with the highest-power-consumption configuration (Fig. \ref{fig4}c).

The experimental validation in a hexagonal PIP mesh further demonstrates that the proposed optimization framework is not restricted to dedicated feedforward architectures, but can also be integrated into multipurpose PIP circuits. This result is particularly relevant because hexagonal meshes constitute one of the most versatile and widely adopted platforms in PIP technology \cite{intro4}, enabling multifunctional and energy-efficient signal processing and computing applications using the same optical hardware.\\

Although the present work focuses on thermo-optic phase shifters, the proposed optimization framework is fundamentally independent of the specific phase-shifter technology because it relies on the geometric interpretation of unitary transformations on the Bloch sphere rather than on the specific physical mechanism responsible for the phase shift. In this context, the key technological aspect determining the implementation of the optimized trajectories is the physical accessibility of both clockwise and counterclockwise $R_{\hat{\mathrm{\textbf{z}}}}$ rotations within each parallel-waveguide section of the BB. Specifically, technologies enabling continuous phase tuning are particularly well suited to the proposed method because they allow continuous control of the $R_{\hat{\mathrm{\textbf{z}}}}$ rotations. 

Table \ref{tab1} summarizes the main characteristics of representative phase-shifter technologies compatible with large-scale SOI-based PIP platforms. In particular, technologies in which the refractive-index variation is intrinsically constrained to a single sign, such as thermo-optic \cite{thermo2} and MEMS \cite{mems3}, generally require two independently actuated phase shifters per $R_{\hat{\mathrm{\textbf{z}}}}$ rotation to physically access both rotation directions on the Bloch sphere. In contrast, technologies enabling both positive and negative refractive-index variations, including free-carrier electro-optic \cite{electro} and ferroelectric BTO-based phase shifters \cite{bto}, can provide bidirectional $R_{\hat{\mathrm{\textbf{z}}}}$ rotations using a single active phase shifter per parallel-waveguide section. However, these technologies may also involve increased insertion losses and larger device footprints, highlighting the broader system-level trade-offs associated with large-scale PIP implementations. 
\clearpage

\begin{table}[h]
\centering
\footnotesize
\renewcommand{\arraystretch}{1.3}
\setlength{\tabcolsep}{3pt}

\begin{tabular}{>{\raggedright\arraybackslash}p{2.3cm}
                >{\raggedright\arraybackslash}p{2.2cm}
                >{\raggedright\arraybackslash}p{1.8cm}
                >{\raggedright\arraybackslash}p{2.3cm}
                >{\raggedright\arraybackslash}p{1.3cm}
                >{\raggedright\arraybackslash}p{1.8cm}
                >{\raggedright\arraybackslash}p{1cm}}
\toprule
\toprule

\textbf{Technology} 
& \textbf{Physical Mechanism} 
& \textbf{Index variation} 
& \textbf{Bidirectional $R_z$ rotation} 
& \textbf{$P_\pi$ (mW)} 
& \textbf{Insertion Losses (dB)} 
& \textbf{Length ($\mu$m)} \\

\midrule
\midrule

Thermo-optic (w/o trenches)\cite{thermo} 
& Temperature dependence
& $\Delta n > 0$
& $2\,\mathrm{PS}/R_{\hat{\mathrm{\textbf{z}}}}$
& 25
& 0.23
& 60 \\
\hline
Thermo-optic (with trenches) \cite{intro13}
& Temperature dependence
& $\Delta n > 0$
& $2\,\mathrm{PS}/R_{\hat{\mathrm{\textbf{z}}}}$
& 1.35
& 0.48
& 220 \\
\hline
MEMS \cite{mems1,mems2}
& Electrostatic actuation
& $\Delta n > 0$
& $2\,\mathrm{PS}/R_{\hat{\mathrm{\textbf{z}}}}$
& 0.001
& 0.33
& 50 \\
\hline
BTO \cite{bto}
& Ferroelectric
& $\Delta n > 0$
$\Delta n < 0$
& $1\,\mathrm{PS}/R_{\hat{\mathrm{\textbf{z}}}}$
& $5.6\cdot10^{-4}$
& 1.48
& 350 \\
\hline
Electro-optic \cite{electro}
& Free-carrier dispersion
& $\Delta n > 0$
$\Delta n < 0$
& $1\,\mathrm{PS}/R_{\hat{\mathrm{\textbf{z}}}}$
& 30
& 4.7
& 500 \\
\hline
\end{tabular}

\caption{Representative phase-shifter technologies for large-scale silicon-on-insulator PIP circuits. The table summarizes the underlying physical mechanism, accessible refractive-index variation ($\Delta n$), implementation of bidirectional $R_{\hat{\mathrm{\textbf{z}}}}$ rotations within each parallel-waveguide section of the BB using one or two active phase shifters, $\pi$-phase shift power ($P_\pi$), insertion losses, and device length. (PS: phase shifter).}
\label{tab1}
\end{table}

While the global phase was omitted throughout the present optimization method, the proposed framework can be straightforwardly extended to scenarios in which the global phase must be preserved, such as coherent-detection architectures \cite{singlechip,complex1,complex2}. In these cases, the global phase associated with each BB can be incorporated through an additional common phase contribution to the output $R_{\hat{\mathrm{\textbf{z}}}}$ rotation. In particular, this additional global phase does not modify the optimization strategy, since it corresponds to a common phase offset that does not require independent optimization, provided that it is restricted to the fundamental interval $[0,2\pi)$. Accordingly, the proposed energy optimization method remains directly applicable to coherent PIP systems.
All in all, the proposed framework provides a general route to designing more energy-efficient large-scale PIP processors, particularly as this system-on-chip technology continues to evolve toward increasingly complex signal processing and computing platforms in both classical and quantum domains.\\

{\Large \textbf{4 Methods}}\\\\
{\textbf{4.1 Manufacturing}}\\

The designed photonic integrated circuit was manufactured by Advanced Micro Foundry, according to a standard SOI process. The chip was fabricated from a SOI wafer with a 220 nm slab thickness, with 500 nm single-mode waveguides defined using deep ultraviolet lithography (193 nm). The waveguide sections with phase shifters (power consumption of 1.35 mW$/\pi$) are obtained by depositing a thin heater layer over the waveguide, made of 120 nm titanium nitride, which will be powered by metal DC tracks of 2000 nm thickness deposited in the final manufacturing stages. The photodetectors were integrated using germanium-on-silicon technology, with responsivities of up to 0.85 A/W. 
\clearpage
\newpage

{\textbf{4.2 Experimental setup}}\\

The experimental setup consists of a tunable continuous-wave (CW) laser (EXFO T100S-HP) followed by a polarization controller. Vertical optical coupling to the chip was achieved through grating couplers using single-mode fibers. Electrical access to the device was provided by electrical probes. At the detection stage, the photocurrents were measured using four independent source-measure units. The thermo-optic phase shifters were electronically controlled using Qontrol programmable power supplies operated through dedicated control software.\\

{\Large \textbf{Data availability}}\\

All data are available from the corresponding author upon request.\\

{\Large \textbf{Code availability}}\\

All codes are available from the corresponding author upon request.\\

{\Large \textbf{Declarations}}\\

The authors declare no conflicts of interest.\\

{\Large \textbf{Acknowledgements}}\\

This work was supported by ERC-ADG-2022-101097092 ANBIT, ERC-POC-2025-1 101241773 TRANSBIT MESH, GVA PROMETEO 2021/015 research excellency award, Fundación BBVA Programa de Investigación Fundamentos 2024 API project, Ministerio de Ciencia y Universidades Plan Complementario de Comunicación Cuántica projects QUANTUMABLE-1 and QUANTUMABLE-2, and HUB de Comunicaciones Cuánticas. \\

{\Large \textbf{Author contributions}}\\

Andrés Macho Ortíz conceived the energy optimization method at the building block level and Pablo Martínez-Carrasco extended the framework across the PIP mesh. Pablo Martínez-Carrasco and José Roberto Rausell carried out the numerical simulations and experimental measurements. Andrés Macho Ortíz  and José Capmany supervised the work. All authors contributed to the preparation of the manuscript. 

\clearpage
\newpage
\printbibliography
\end{refsection}

\clearpage
\newpage
\pagenumbering{arabic}
\setcounter{page}{1}
\setcounter{tocdepth}{2}

\title{Energy-efficient programmable integrated photonics \\via optimized Euler rotations.\\
Supplementary Information}

\date{}
\maketitle
\begin{center}

{\large
Pablo Martínez-Carrasco Romero,$^{a,*}$ 
Andrés Macho-Ortiz,$^{a,*}$ \\
José Roberto Rausell-Campo,$^{b}$ 
Francisco Javier Fraile-Peláez,$^{b}$ 
and José Capmany$^{a,c}$
}

\vspace{0.5cm}

\small
$^{a}$iTEAM Research Institute, Universitat Politècnica de València, Valencia, 46022, Spain \\

$^{b}$Dept. Teoría de la Señal y Comunicaciones, Universidad de Vigo E.I. Telecomunicación,\\
Campus Universitario, E-36202 Vigo (Pontevedra), Spain \\

$^{c}$iPronics, Programmable Photonics, S.L, Valencia, 46022, Spain \\

\vspace{0.3cm}

*Corresponding authors: 
pmarrom@iteam.upv.es, amachor@iteam.upv.es

\end{center}
\tableofcontents
\clearpage
\newpage
\begin{refsection}

\section{Supplementary 1: Total phase accumulation metric}
\label{supp1}

In the main text, the total phase accumulation $|\alpha| + |\beta| + |\gamma|$ was introduced as a physically meaningful metric associated with the energy cost of implementing a given 2×2 unitary transformation. In this section, we discuss the relation between this quantity and the energetic cost of representative phase-shifter technologies compatible with large-scale SOI-based PIP platforms.

In \textit{thermo-optic} phase shifters, the induced phase variation is approximately proportional to the electrical power dissipated by the heater under steady-state operation, since the refractive-index variation scales linearly with temperature over the operating range \cite{thermo2},\cite{ribeiro2020column}. In this way, the total phase accumulation provides a direct metric for the overall electrical power consumption of the 2×2 unitary PIP processor.

In electrostatically actuated \textit{MEMS} phase shifters, the phase shift is controlled through a voltage-induced mechanical displacement \cite{mems3},\cite{reed2014recent}. Unlike thermo-optic heaters, the static power required to hold a given phase can be negligible, and the dominant energetic cost is associated with charging the actuator capacitance and switching between configurations. Therefore, the total phase accumulation should be interpreted as a metric for the required actuation range or switching energy, rather than as a quantity universally proportional to static power consumption.

In \textit{free-carrier}-based phase shifters, the optical phase shift is induced through electrically controlled variations of the carrier concentration, which modify the effective refractive index of the waveguide \cite{electro},\cite{reed2014recent}. Depending on the specific implementation (carrier injection or depletion), the associated energetic cost may arise from static electrical dissipation, capacitive charging, or dynamic switching processes. Consequently, the total phase accumulation provides a metric for the required electrical actuation, although its precise relation to power consumption depends on the phase-shifter architecture and operating regime.

In \textit{ferroelectric} electro-optic phase shifters, such as barium titanate (BTO)-based devices integrated on silicon-on-insulator (SOI) platforms, the phase variation is induced through electrically controlled refractive-index modulation arising from the electro-optic effect \cite{geler2022ferroelectric},\cite{abel2019large}. In these devices, the static power consumption can be extremely low, and the dominant energetic cost is typically associated with capacitive charging and electrical control. Accordingly, the total phase accumulation should be interpreted as a metric for the required electro-optic actuation rather than as a quantity universally proportional to static power dissipation.
\clearpage
\newpage

\section{Supplementary 2: Energy optimization of 2x2 unitary matrices}
\label{supp2}

In this section we further develop the analysis by deriving the different angle decompositions associated with a given 2x2 unitary transformation and comparing their corresponding total phase accumulations. To do so, we first formally derive the transfer matrices of the optical components composing the BBs and establish the relationship between the implemented phase shifts and their equivalent rotation angles. The transfer matrix of a parallel-waveguide section incorporating two independently actuated phase shifters — introducing phase shifts of $\alpha_1$ and $\alpha_2$  rad, respectively — is given by:

\begin{equation}
\begin{aligned}
    T_{\mathrm{PS}}(\alpha_{1}, \alpha_{2}) &= 
    \begin{pmatrix} 
    e^{i\alpha_{1}} & 0\\
    0 & e^{i\alpha_{2/2}}
    \end{pmatrix} =
    e^{i\left(\alpha_{2}-\alpha_{1}\right)/2} 
    \begin{pmatrix} 
    e^{-i\left(\alpha_{2}-\alpha_{1}\right)/2} & 0\\
    0 & e^{i\left(\alpha_{2}-\alpha_{1}\right)/2}
    \end{pmatrix}\\
    &\equiv e^{i\left(\alpha_{2}-\alpha_{1}\right)/2} R_{\hat{\mathrm{\textbf{z}}}}(\alpha = \alpha_{2}-\alpha_{1})
\end{aligned}
\end{equation}

This expression shows that the relative phase difference between the two waveguides implements a tunable rotation around the z-axis of the Bloch sphere, whereas the common phase term corresponds to a (non-observable) global phase contribution \cite{macho2021optical}. Similarly we can compare the transfer matrix of the MMI with an ideal rotation matrix, in this case the comparison is direct: 

\begin{equation}
    T_{\mathrm{MMI}} =  \dfrac{1}{\sqrt{2}}
    \begin{pmatrix} 
    1 & i\\
    i & 1
    \end{pmatrix}
    \equiv R_{\hat{\mathrm{\textbf{x}}}}(-\pi/2) 
\end{equation}

Based on this, the overall transfer function of the 2x2 unitary processor, excluding the common phase term introduced by the accumulated phase shifters, can be expressed as:

\begin{equation}
\begin{aligned}
    &T_{2\times2}(\alpha_{1,2},\beta_{1,2},\gamma_{1,2}) = T_{\mathrm{PS}}(\gamma_{1,2})T_{\mathrm{MMI}}T_{\mathrm{PS}}(\beta_{1,2})T_{\mathrm{MMI}}T_{\mathrm{PS}}(\alpha_{1,2}) =\\
    &=i
    \begin{pmatrix} 
     -e^{-i[(\alpha_{2}-\alpha_{1})+(\gamma_{2}-\gamma_{1})]/2}\sin\left(\dfrac{\beta_{2}-\beta_{1}}{2}\right) & e^{i[(\alpha_{2}-\alpha_{1})-(\gamma_{2}-\gamma_{1})]/2}\cos\left(\dfrac{\beta_{2}-\beta_{1}}{2}\right) \\
     e^{-i[(\alpha_{2}-\alpha_{1})-(\gamma_{2}-\gamma_{1})]/2}\cos\left(\dfrac{\beta_{2}-\beta_{1}}{2}\right) & e^{i[(\alpha_{2}-\alpha_{1})+(\gamma_{2}-\gamma_{1})]/2}\sin\left(\dfrac{\beta_{2}-\beta_{1}}{2}\right) \\
    \end{pmatrix}
\end{aligned}
\end{equation}

We can then relate the angles of the physical implementation to the corresponding angles of the overall transformation expressed solely in terms of rotation matrices:

\begin{equation}
\begin{aligned}
    T_{2\times2}(\alpha,\beta,\gamma) &= R_{\hat{\mathrm{\textbf{z}}}}(\gamma)R_{\hat{\mathrm{\textbf{x}}}}(-\pi/2)R_{\hat{\mathrm{\textbf{z}}}}(\beta)R_{\hat{\mathrm{\textbf{x}}}}(-\pi/2)R_{\hat{\mathrm{\textbf{z}}}}(\alpha)  =  \\ &=i
    \begin{pmatrix} 
     -e^{-i(\alpha+\gamma)/2}\sin\left(\dfrac{\beta}{2}\right) & e^{i(\alpha-\gamma)/2}\cos\left(\dfrac{\beta}{2}\right) \\
     e^{-i(\alpha-\gamma)/2}\cos\left(\dfrac{\beta}{2}\right) & e^{i(\alpha+\gamma)/2}\sin\left(\dfrac{\beta}{2}\right) \\
    \end{pmatrix}
\end{aligned}
\end{equation}

The resulting relationships between the physical phase shifts and the corresponding theoretical angles are:

\begin{equation}
     \alpha = \alpha_{2}-\alpha_{1}
\end{equation}
\begin{equation}
     \beta = \beta_{2}-\beta_{1}
\end{equation}
\begin{equation}
     \gamma = \gamma_{2}-\gamma_{1}
\end{equation}

Having access to both phase shifters allows for the implementation of both positive and negative angles for $\alpha$, $\beta$ and $\gamma$.

For matrix decomposition and implementation, the global phase factor is disregarded. Therefore, we consider a unitary matrix $U$ belonging to SU(2), a Lie subgroup of U(2) with $det(U) = 1$

\begin{equation}
    U = 
    \begin{pmatrix} 
     U_{11} & U_{12} \\
     U_{21} & U_{22} \\
    \end{pmatrix}
    = 
    \begin{pmatrix} 
     a & b \\
     -b^{*} & a^{*} \\
    \end{pmatrix}
\end{equation}

In the case of a decomposition of unitary matrices in SU(2), the use of complex algorithms is not necessary since the matrix dimension is very small, allowing a term-by-term correspondence between the desired matrix to implement and the device's transfer function, $U = T_{2\times2}(\alpha,\beta,\gamma)$.

\begin{equation}
     U_{11} = a = -ie^{-i(\alpha+\gamma)/2}\sin\left(\dfrac{\beta}{2}\right)
\end{equation}
\begin{equation}
     U_{12} = b= ie^{i(\alpha-\gamma)/2}\cos\left(\dfrac{\beta}{2}\right) 
\end{equation}
\begin{equation}
     U_{21} = -b^{*}= ie^{-i(\alpha-\gamma)/2}\cos\left(\dfrac{\beta}{2}\right) 
\end{equation}
\begin{equation}
     U_{22} = a^{*}= ie^{i(\alpha+\gamma)/2}\sin\left(\dfrac{\beta}{2}\right)
\end{equation}

Among these relations, only two of the four can be used to determine the angle values, as the remaining two are merely re-expressions of the others and offer no additional information. Using only two equations to solve for three angles results in an overdetermined system, allowing for multiple possible solutions. There are several ways to solve for the angles, but they all yield the same results. The last angle to be determined depends on the first two, depending on the order in which the angles are solved. For the purpose of illustrating the process, in this work we will use the first two equations to determine the angles.

\begin{equation}
     \dfrac{U_{11}}{U_{12}} = \dfrac{a}{b}=  -e^{-i\alpha}\tan\left(\dfrac{\beta}{2}\right)
\end{equation}

We find the possible solutions for $\beta$ in $[-\pi, \pi)$:

\begin{equation}
     \left|\dfrac{U_{11}}{U_{12}}\right| =  \left|\tan\left(\dfrac{\beta}{2}\right)\right|
\end{equation}

\begin{equation}
     \beta=  \pm2\arctan\left(\left|\dfrac{U_{11}}{U_{12}}\right|\right) 
\end{equation}

The two possible solutions for $\beta$ (one with positive sign and the other one with negative sign) yield two distinct sets of $\alpha,\beta,\gamma$. values. From each set, the corresponding two possible values of $\alpha$ and $\gamma$ can be determined as shown below (these solutions can be obtained in different ways).

\begin{equation}
    \alpha = -\mathrm{Arg}\left(\dfrac{-U_{11}}{U_{12}\tan\left(\dfrac{\beta}{2}\right)}\right)
\end{equation}

\begin{equation}
    \gamma = -2\mathrm{Arg}\left(\dfrac{iU_{11}e^{i\alpha/2}}{\sin\left(\dfrac{\beta}{2}\right)}\right)
\end{equation}

where Arg denotes the principal value of the complex argument, restricted to the fundamental interval [0,$2\pi$). The two sets of angles obtained include a mix of positive and negative values. Negative angles cannot be directly implemented in photonic circuits with a single active phase shifter; however, their complementary angles can be realized at the cost of increased energy consumption. This is possible because $R_{\hat{\mathrm{\textbf{z}}}}(\varphi + 2\pi) = R_{\hat{\mathrm{\textbf{z}}}}(2\pi)R_{\hat{\mathrm{\textbf{z}}}}(\varphi) = -R_{\hat{\mathrm{\textbf{z}}}}(\varphi)$, which corresponds only to a global phase shift for the entire transfer matrix. Depending on the set of angles, two possible implementations with all-positive angles can be obtained. One of them requires a larger total angular excursion than the other, and both are only applicable to systems with a single active phase shifter. In contrast, a system with both phase shifters active can select either of these solutions or those involving negative angles, choosing the option that minimizes the overall angular excursion.

\clearpage
\newpage

\section{Supplementary 3: Energy optimization of NxN unitary matrices}
\label{supp3}

This section provides a more detailed explanation of the extrapolation of the energy optimization process based on rotations in higher-dimensional matrices. Any NxN unitary matrix in integrated photonics can be implemented by decomposing it into a sequence of 2x2 transformations using beam splitters and phase shifters, followed by a diagonal phase-shifter layer at the output, as in both Reck- and Clements-type architectures \cite{reck},\cite{clements}. These transformations are related to a fundamental building block described by the transfer matrix $T_{\mathrm{BB}}(\theta_{1,2}, \phi_{1,2})$ (assuming all phase shifters are active), given by:

\begin{equation}
    T_{\mathrm{BB}}(\theta_{1,2}, \phi_{1,2}) = ie^{i(\phi_{1}+\phi_{2}+\theta_{1}+\theta_{2})/2}
    \begin{pmatrix} 
    -e^{-i(\phi_{2}-\phi_{1})/2}\sin{(\dfrac{\theta_{2}-\theta_{1}}{2})} & e^{i(\phi_{2}-\phi_{1})/2}\cos{(\dfrac{\theta_{2}-\theta_{1}}{2})}\\
    e^{-i(\phi_{2}-\phi_{1})/2}\cos{(\dfrac{\theta_{2}-\theta_{1}}{2})} & e^{i(\phi_{2}-\phi_{1})/2}\sin{(\dfrac{\theta_{2}-\theta_{1}}{2})}
    \end{pmatrix}
\end{equation}

This block is analogous to the 2×2 matrices $\tilde{T}_{m,n}$ embedded within the two-level unitary matrices $T_{m,n}$, which are used to decompose the $N \times N$ matrix by eliminating its off-diagonal elements \cite{clements}:

\begin{equation}
\begin{aligned}
    \tilde{T}_{m,n}(\theta, \phi) &= R_{\hat{\mathrm{\textbf{x}}}}(-\pi/2)R_{\hat{\mathrm{\textbf{z}}}}(\theta)R_{\hat{\mathrm{\textbf{x}}}}(-\pi/2)R_{\hat{\mathrm{\textbf{z}}}}(\phi)=\\
    &=i\begin{pmatrix} 
    -e^{-i\phi/2}\sin{(\theta/2)} & e^{i\phi/2}\cos{(\theta/2)}\\
     e^{-i\phi/2}\cos{(\theta/2)} & e^{i\phi/2}\sin{(\theta/2)}
    \end{pmatrix}
\end{aligned}
\end{equation}

Thus, the correspondence between the angles of the unitary rotations and their physical implementation through phase shifters is established.

\begin{equation}
\theta = \theta_{2}-\theta_{1}
\end{equation}
\begin{equation}
\phi = \phi_{2}-\phi_{1}
\end{equation}

As in the SU(2) implementation (Supplementary Note \ref{supp2}), each unitary can be optimized to minimize the total rotation angle. In higher dimensions, however, no closed-form relation links the target matrix to the angles of each BB, so a decomposition algorithm is used to optimize them sequentially and reduce the overall angular cost.

In the algorithm, the goal is to eliminate the off-diagonal elements of the matrix $U$ by sequentially applying a series of BB transformations $\tilde{T}_{m,n}$ between optical channels $m$ and $n$. After the decompostion the resulting diagonal left matrix and the $\tilde{T}_{m,n}$ operators allow the reconstuction of the original $U$ matrix. In this cancellation process, the $\arctan$ function appears again yielding two distinct values of $\theta$, one positive and one negative, that affect the two possible solution sets. As a result, two different angle combinations arise for each BB, each associated with a different angular cost. When only a single phase shifter is active, the same combinations are obtained, but expressed in terms of complementary angles.
Algorithm 1 presents the decomposition procedure, which is an extended version of the widely used Clements decomposition algorithm. It incorporates an additional step that compares the total rotation angles before selecting the set to apply in each BB.

\clearpage

\begin{algorithm}
\caption{Unitary Matrix Decomposition Algorithm}
\begin{algorithmic}
\State \textbf{DECOMPOSE {$U(N)$}}
\For{$i=1$ to $N-1$}
    \If {$i$ is odd}
        \For {$j = 0$ to $i-1$}
            \State Find $\{(\theta_{1},\phi_{1}),(\theta_{2},\phi_{2})\}$ such that $\tilde{T}_{(i-j,\;i-j+1)}(\theta,\phi)^{-1}$ nullifies $U_{(N-j,\; i-j)}$
            \If {\ensuremath{|\theta_{1}|+|\phi_{1}| < |\theta_{2}|+|\phi_{2}|}}
                \State Set $\tilde{T}_{(i-j,\;i-j+1)}(\theta_{\min},\phi_{\min}) = \tilde{T}_{(i-j,\;i-j+1)}(\theta_{1},\phi_{1})$
            \Else
                \State Set $\tilde{T}_{(i-j,\;i-j+1)}(\theta_{\min},\phi_{\min}) = \tilde{T}_{(i-j,\;i-j+1)}(\theta_{2},\phi_{2})$
            \EndIf
            \State Update $U = U\,\tilde{T}_{(i-j,\;i-j+1)}(\theta_{\min},\phi_{\min})^{-1}$
        \EndFor
    \Else
        \For {$j = 1$ to $i$}
            \State Find $\{(\theta_{1},\phi_{1}),(\theta_{2},\phi_{2})\}$ such that $\tilde{T}_{(N+j-i-1,\;N+j-i)}(\theta,\phi)$ nullifies $U_{(N+j-i,\; j)}$
            \If {$|\theta_{1}|+|\phi_{1}| < |\theta_{2}|+|\phi_{2}|$}
                \State Set $\tilde{T}_{(N+j-i-1,\;N+j-i)}(\theta_{\min},\phi_{\min}) = \tilde{T}_{(N+j-i-1,\;N+j-i)}(\theta_{1},\phi_{1})$
            \Else
                \State Set $\tilde{T}_{(N+j-i-1,\;N+j-i)}(\theta_{\min},\phi_{\min}) = \tilde{T}_{(N+j-i-1,\;N+j-i)}(\theta_{2},\phi_{2})$
            \EndIf
            \State Update $U = \tilde{T}_{(N+j-i-1,\;N+j-i)}(\theta_{\min},\phi_{\min})\,U$
        \EndFor
    \EndIf
\EndFor
\end{algorithmic}
\label{alg:clements}
\end{algorithm}

BB operations are independent, so optimizing each block and choosing among its angle sets does not affect the others; however, these choices modify the accumulated global phase, which must be implemented by the diagonal phase shifters at the circuit output. For low-dimensional matrices, this can make the diagonal phase shifters consume more than the savings from BB optimization; however, for larger matrices, their contribution becomes negligible compared with the reduction achieved by optimizing the BBs.

\begin{figure}[ht]
\centering\includegraphics[width=0.7\linewidth]{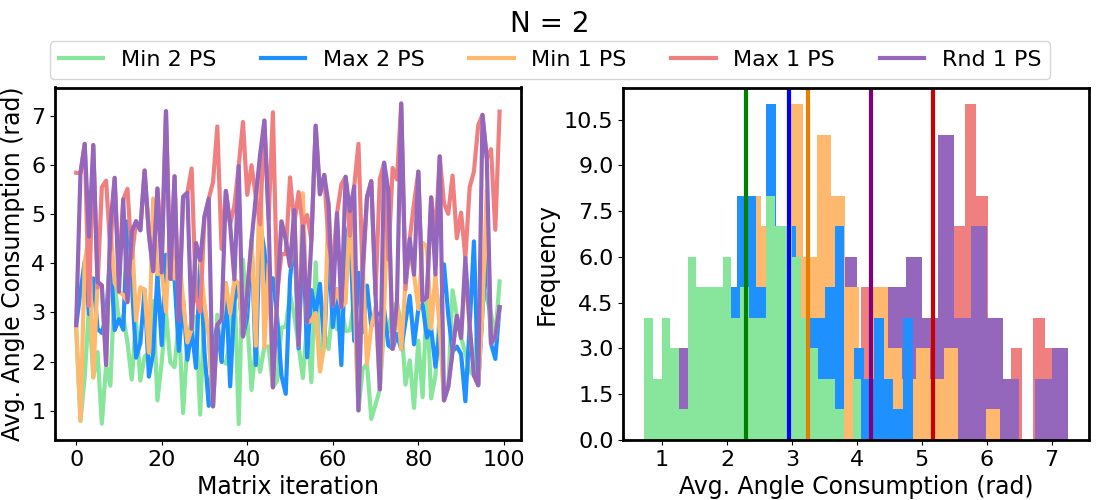}\\
\centering\includegraphics[width=0.7\linewidth]{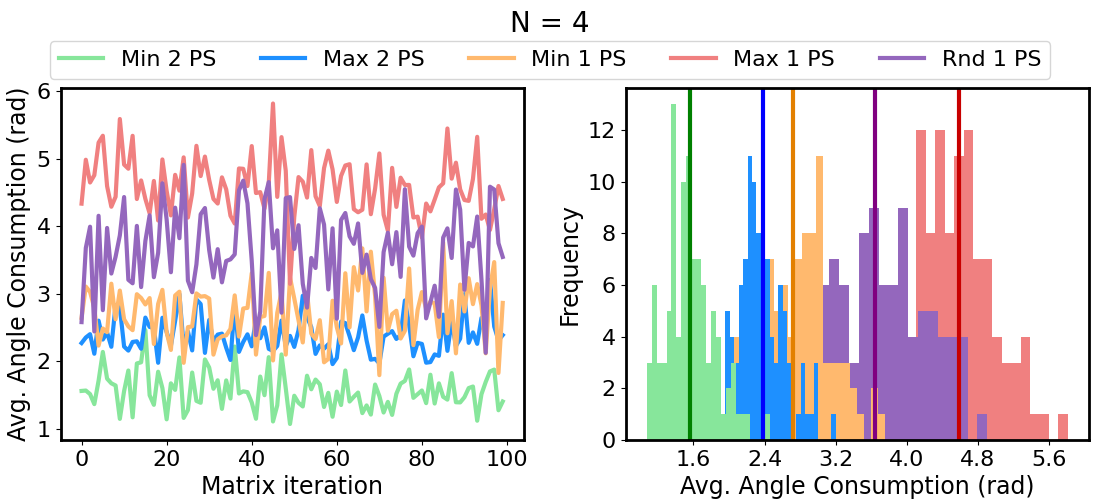}\\ 
\centering\includegraphics[width=0.7\linewidth]{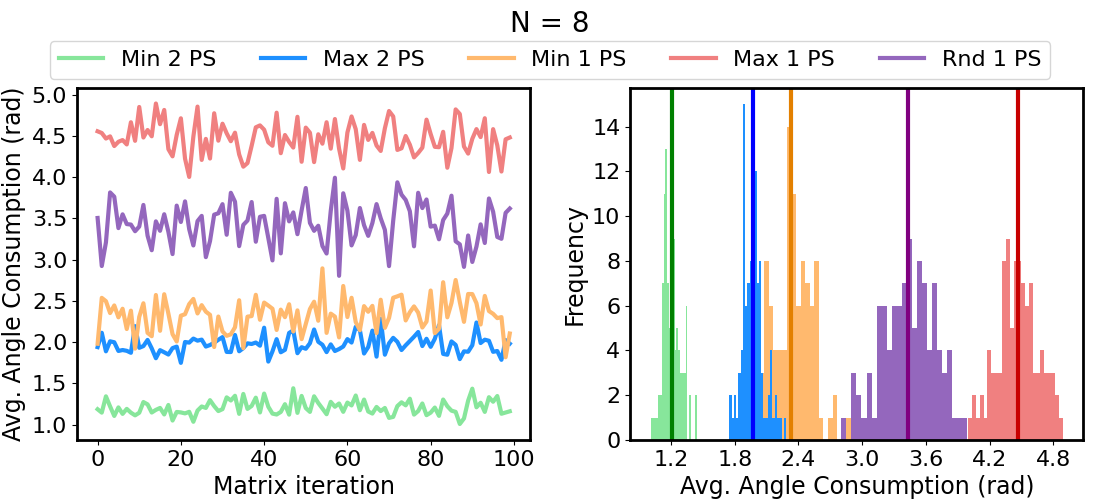}\\
\centering\includegraphics[width=0.7\linewidth]{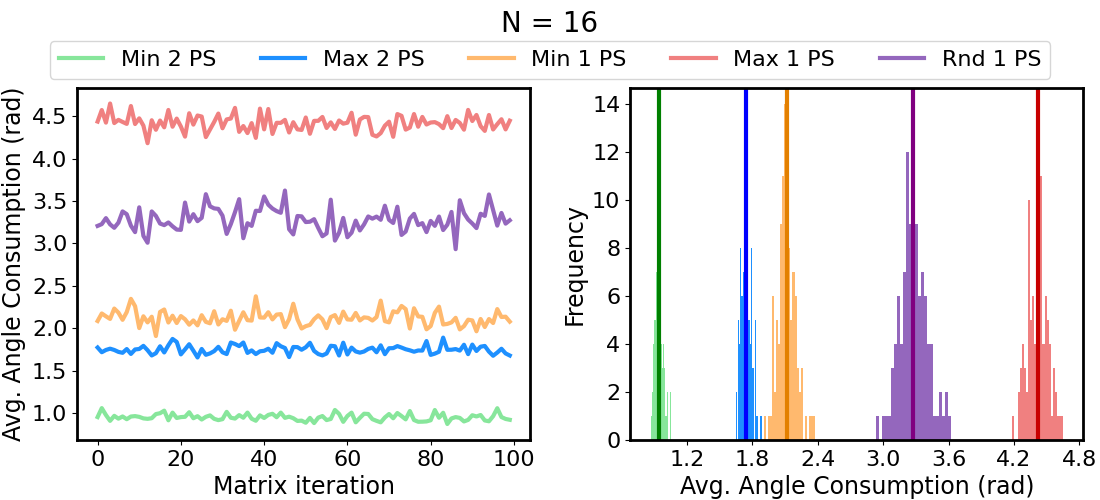}
\caption{Comparison of angular consumption per phase shifter in the decompositions as a function of matrix size (N = 2, 4, 8, 16) and the number of optimized phase shifters.}
\label{fig:supp1}
\end{figure}

We simulated the decomposition of matrices of various sizes using Clements-type architectures, following the same color scheme as in the Main Document. In addition, a purple line shows a random selection of angles for each BB (for a single phase shifter), which typically yields an intermediate consumption between the minimum and maximum. Figure S.\ref{fig:supp1} shows the results for 100 matrices of various sizes (N = 2, 4, 8, 16), where the x-axis represents each of the 100 decomposed matrix elements, and the y-axis indicates the average angular consumption per phase shifter for the different optimization strategies.

\clearpage

The results indicate that for low-dimensional matrices, optimizing the BBs alone does not always achieve the most optimal expected outcome, since, as noted earlier, the BB configuration affects the final diagonal, which still contributes significant angular consumption for small matrices. However, as the matrix dimension increases, the consumption curves diverge, clearly showing that BB optimization leads to overall angular consumption that aligns with theoretical predictions

\clearpage
\newpage

\section{Supplementary 4: Energy optimization of hexagonal meshes}
\label{supp4}

Optical matrices can be implemented in reconfigurable hexagonal meshes by configuring the MZIs to operate either as ideal power splitters without additional phase shifts or as ideal phase shifters \cite{rausell2025programming}. This functionality requires prior calibration of all phase shifters in the MZIs that constitute the mesh. The transfer matrix of an MZI in the hexagonal mesh, when both internal phase shifters are active, $\beta_{1}$ and $\beta_{2}$, is given by:

\begin{equation}
    T_{\mathrm{MZI}}(\beta_{1,2}) =ie^{i(\beta_{1}+\beta_{2})/2}
    \begin{pmatrix} 
     -\sin\left(\dfrac{\beta_{2}-\beta_{1}}{2}\right) & \cos\left(\dfrac{\beta_{2}-\beta_{1}}{2}\right) \\
     \cos\left(\dfrac{\beta_{2}-\beta_{1}}{2}\right) & \sin\left(\dfrac{\beta_{2}-\beta_{1}}{2}\right) \\
    \end{pmatrix}
\end{equation}

The final goal is to obtain a structure formed by concatenated MZIs that reproduces the behavior of the ideal BB of feedforward matrix architectures with two active phase shifters, thereby enabling the implementation of both positive and negative angles:

\begin{equation}
    T_{\mathrm{BB}}(\theta_{1,2}, \phi_{1,2}) = ie^{i(\phi_{1}+\phi_{2}+\theta_{1}+\theta_{2})/2}
    \begin{pmatrix} 
    -e^{-i(\phi_{2}-\phi_{1})/2}\sin{(\dfrac{\theta_{2}-\theta_{1}}{2})} & e^{i(\phi_{2}-\phi_{1})/2}\cos{(\dfrac{\theta_{2}-\theta_{1}}{2})}\\
    e^{-i(\phi_{2}-\phi_{1})/2}\cos{(\dfrac{\theta_{2}-\theta_{1}}{2})} & e^{i(\phi_{2}-\phi_{1})/2}\sin{(\dfrac{\theta_{2}-\theta_{1}}{2})}
    \end{pmatrix}
\end{equation}

The resulting matrices are very similar in terms of coupling control and only require allocating a small number of additional mesh resources to operate as external phase shifters. To this end, the MZIs located before the coupling-control MZI are configured to operate in a perfect cross state, allowing all the optical power to pass through while still enabling control of the total accumulated phase along the path, as illustrated below:

\begin{equation}
    T_{\mathrm{PS}}(\beta_{1,2}) =ie^{i(\beta_{1}+\beta_{2})/2}
    \begin{bmatrix} 
     0 & 1 \\
     1 & 0 \\
    \end{bmatrix}
\end{equation}

To this end, the following two equations must be applied (for the case in which the final phase shifter is implemented in the upper arm of the BB, $\phi_{1}$):

\begin{equation}
    \beta_{1}+\beta_{2}=\phi_{1} 
\end{equation}
\begin{equation}
    \beta_{1}-\beta_{2}=0
\end{equation}
\clearpage

\begin{figure}[ht]
\centering\includegraphics[width=0.3\linewidth]{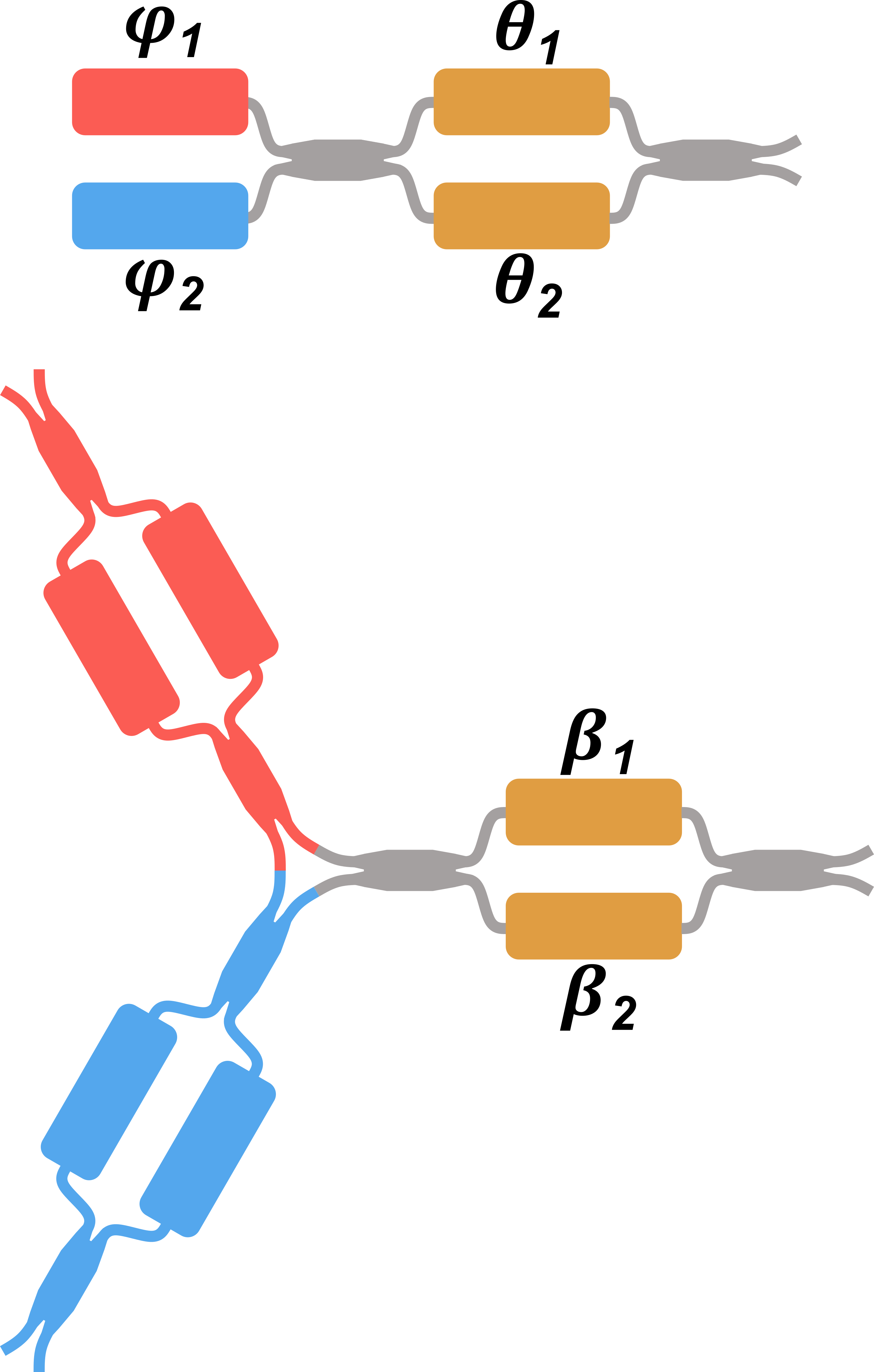} 
\caption{Comparison between the standard building block, consisting of an MZI with two internal and two external phase shifters enabling the implementation of both positive and negative angles, and its equivalent realization in a hexagonal mesh.}
\label{fig:supp2}
\end{figure}

Figure S.\ref{fig:supp2} illustrates the relationship between BB implementation for matrix decomposition using a feedforward network and a hexagonal mesh architecture.

\begin{figure}[ht]
\centering\includegraphics[width=0.8\linewidth]{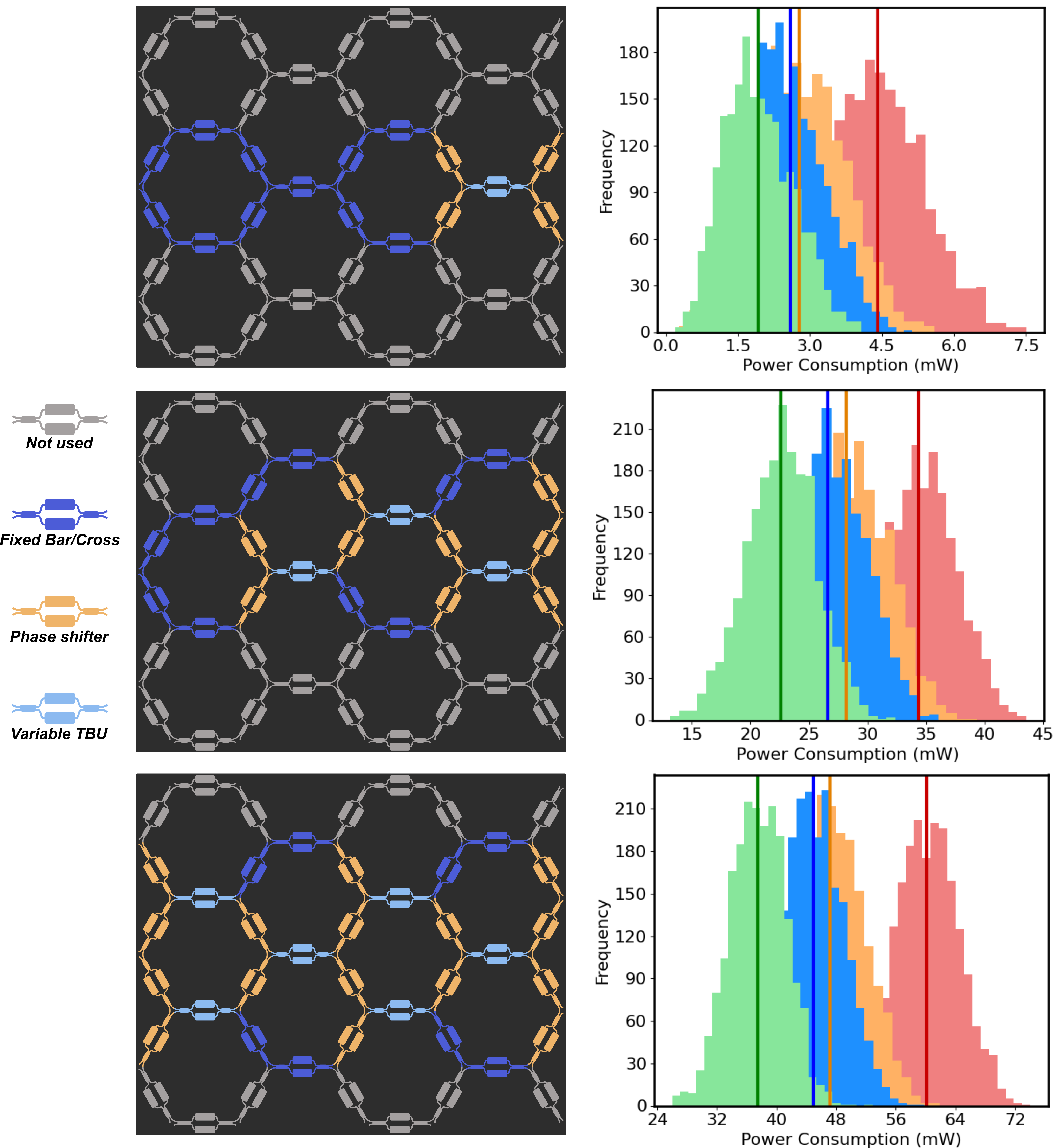} 
\caption{Power consumption for implementing random matrices of varying sizes on a hexagonal mesh.
}
\label{fig:supp3}
\end{figure}

Figure S.\ref{fig:supp3} presents the measurement results for implementing arbitrary unitary matrices on a hexagonal mesh for 2×2, 3×3, and 4×4 cases. It shows that the improvement in power consumption becomes more significant as the matrix dimension increases, due to the different implementation strategies considered.

\clearpage
\newpage
\printbibliography
\end{refsection}

\end{document}